\documentclass[journal]{IEEEtran}

\usepackage{graphicx}
\usepackage[T1]{fontenc}
\usepackage{amsmath}
\usepackage[cmintegrals]{newtxmath}
\usepackage{url,multirow,morefloats,floatflt,cancel,tfrupee}
\usepackage{colortbl}
\usepackage{xcolor}
\usepackage{pifont}
\usepackage[nointegrals]{wasysym}
\begin{document}

\title{Coincidence-pumping upconversion detector based on passively synchronized fiber laser system}

\author{
	Weiyan Kang, Bowen Li, Yan Liang, Qiang Hao, Ming Yan, Kun Huang, and Heping Zeng
	\thanks{Manuscript received $\times \times$; revised $\times \times$; accepted $\times \times$. Date of publication $\times \times$; date of current version $\times \times$. This work was supported by the Program for Professor of Special Appointment (Eastern Scholar) at Shanghai Institutions of Higher Learning, Science and Technology Innovation Program of Basic Science Foundation of Shanghai under Grant 18JC1412000. \textit{(Corresponding author: Kun Huang.)}}
        \thanks{W. Kang, B. Li, Y. Liang, Q. Hao, K. Huang, and H. Zeng are with Shanghai Key Laboratory of Modern Optical System, and Engineering Research Center of Optical Instrument and System, Ministry of Education, School of Optical Electrical and Computer Engineering, University of Shanghai for Science and Technology, Shanghai 200093, China (e-mail: khuang@lps.ecnu.edu.cn).}
        \thanks{M. Yan, K. Huang, and H. Zeng are with State Key Laboratory of Precision Spectroscopy, East China Normal University, Shanghai 200062, China.}
        \thanks{H. Zeng is with Jinan Institute of Quantum Technology, Jinan, Shandong 250101, China.}
        \thanks{Color versions of one or more of the figures in this letter are available online at http://ieeexplore.ieee.org.}
        \thanks{Digital Object Identifier $\times \times \times$}
        }

\maketitle

\begin{abstract}
We experimentally demonstrated a high-performance frequency upconversion detector for telecom-band photons based on a passively synchronized fiber laser system. The involved coincidence pumping technique enabled to spectrally convert the pulsed infrared photons into the visible regime with a conversion efficiency of 72\%. The overall detection efficiency of the upconversion detector reached to 30\% with a low noise equivalent power of $3\times10^{-17}\ \text{W/Hz}^{1/2}$. In contrast to previous demonstrations, the whole upconversion detection system was constructed in an all-polarization-maintaining fiber structure, thus favoring substantial improvement of compactness and robustness. Moreover, the long-term stability was manifested by at least ten-hour operation with a relative fluctuation of count rates as small as 0.26\%. The achieved features here would be desirable in many practical applications requiring efficient and robust coherent manipulation of pulsed optical fields by nonlinear frequency conversion.
\end{abstract}

\IEEEpeerreviewmaketitle

\section{Introduction}
Frequency upconversion detection provides an attractive photon-counting technique based on optical nonlinear frequency conversion \cite{Ma2012PR}. The underlying process can be exemplified with the sum frequency generation (SFG), where an input field is mixed with an intensive pump in quadratic nonlinear medium to generate an output field with the sum of the input frequencies \cite{Xiang2018PRA}. The resulting upconverted field typically in the visible regime is then detected by well-developed silicon avalanche photodiodes (Si-APDs) \cite{Hadfield2009NP}. The upconversion detector could thus offer an effective way to extend operation wavelengths of high-performance silicon detectors into spectral regions usually hard to access, especially for infrared bands \cite{Albota2004OL, Pan2006OL, Kamada2008OL, Ma2018JOSAB}. Although the detection of telecom-band single photons can be readily realized by InGaAs/InP APDs and superconducting single-photon detectors, yet the upconversion single-photon detector exhibits comprehensive competence by inheriting intrinsic properties of Si-based detectors, such as high efficiency, low dark noise, room-temperature operation and relatively low cost \cite{Hadfield2009NP}. Moreover, the combination with the Si-based electron multiplying CCD has enabled to demonstrate high-resolution infrared imaging at the few-photon level \cite{Huang2012APL, Dam2012NP}. Additionally, due to the phase-matching requirement in the frequency conversion process, the upconversion detectors could offer some unique features such as tunable wavelength acceptance \cite{Thew2008APL} and polarization sensitivity \cite{Kaiser2019OE}, hence rendering them useful in all-optical nonlinear signal processing. 

Particularly, the so-called coincidence-pumping upconversion detector has attracted increasing attention over the past decade \cite{Vandevender2004JMO, Gu2010APL, Tang2015PTL}, where the signal photons are synchronously gated by pulsed pump. Consequently, the pulsed pumping scheme could leverage the high peak power and ultrashort excitation window, which would help to increase the conversion efficiency and reduce the background noise \cite{Vandevender2004JMO, Gu2010APL}. Such configuration is especially suitable to detect infrared signal photons within the regularly pulsed envelope, for example in applications such as long-distance telecommunication, ultra-sensitive pump-probe sensing and low-light active infrared imaging. Compared with solid-state lasers, fiber lasers could offer advantages like compactness, easy maintenance and free of optics alignment. Further optimization of detector performance has been investigated by temporally and spectrally controlling the involved interacting optical fields from two passively synchronized fiber lasers \cite{Gu2010APL}. However, the demonstrated dual-color fiber laser synchronization system was implemented in a non-polarization-maintaining structure (non-PM), which was thus inevitably susceptible to ambient perturbations. Consequently, self-starting and long-term operation of the synchronously pumping upconversion detector would be hard to obtain in practice. 

In this paper, we devised and implemented a coincidence pumping upconversion detector based on an all-polarization-maintaining fiber laser synchronization system, where the infrared signal photons at 1.58 $\mu$m were tightly gated by the passively synchronized pump pulses at 1.03 $\mu$m. Thanks to the spectro-temporal engineering of the involved optical pulses, an overall peak detection efficiency of 30\%, corresponding to a maximum conversion efficiency of 72\%, was achieved with a noise equivalent power of $3\times10^{-17}\ \text{W/Hz}^{1/2}$. Moreover, the long-term stability was manifested by at least ten-hour operation with a relative fluctuation of count rates as small as 0.26\%. The robust and efficient upconversion detector demonstrated here would find potential applications including remote spectroscopic detection and ultra-sensitive infrared imaging.

\section{Experimental setup}

\begin{figure}[t!]
\centering
\includegraphics[width=1\columnwidth]{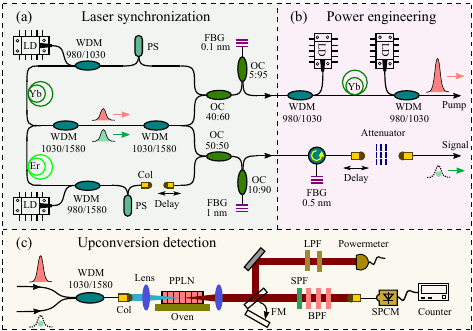}
\caption{(a) All-PM synchronization system for Yb- and Er-doped fiber lasers (YDFL and EDFL). The synchronized mode-locking operation is realized by nonlinear cross-phase modulation between dual-color pulses within the common single-mode fiber. (b) Output pulses from YDFL and EDFL are then properly scaled in power to provide pump and signal sources by amplification and attenuation, respectively. (c) Single-photon upconversion detection based on passively synchronous pumping. LD: Laser diode; WDM: wavelength division multiplexer; Yb: ytterbium-doped gain fiber; Er: erbium-doped gain fiber; OC: output coupler; PS: phase shifter;  FBG: fiber Bragg grating; Col: collimator; FM: flip mirror; PPLN: periodically-poled lithium niobate crystal; LPF, SFP and BPF: long-, short- and band-pass optical filter; SPCM: single photon counting module. Note that the WDMs were operated in a reflective fashion for the pump injection into gain fibers.}
\label{fig1}
\end{figure}

\begin{figure}[b!]
\centering
\includegraphics[width=1\columnwidth]{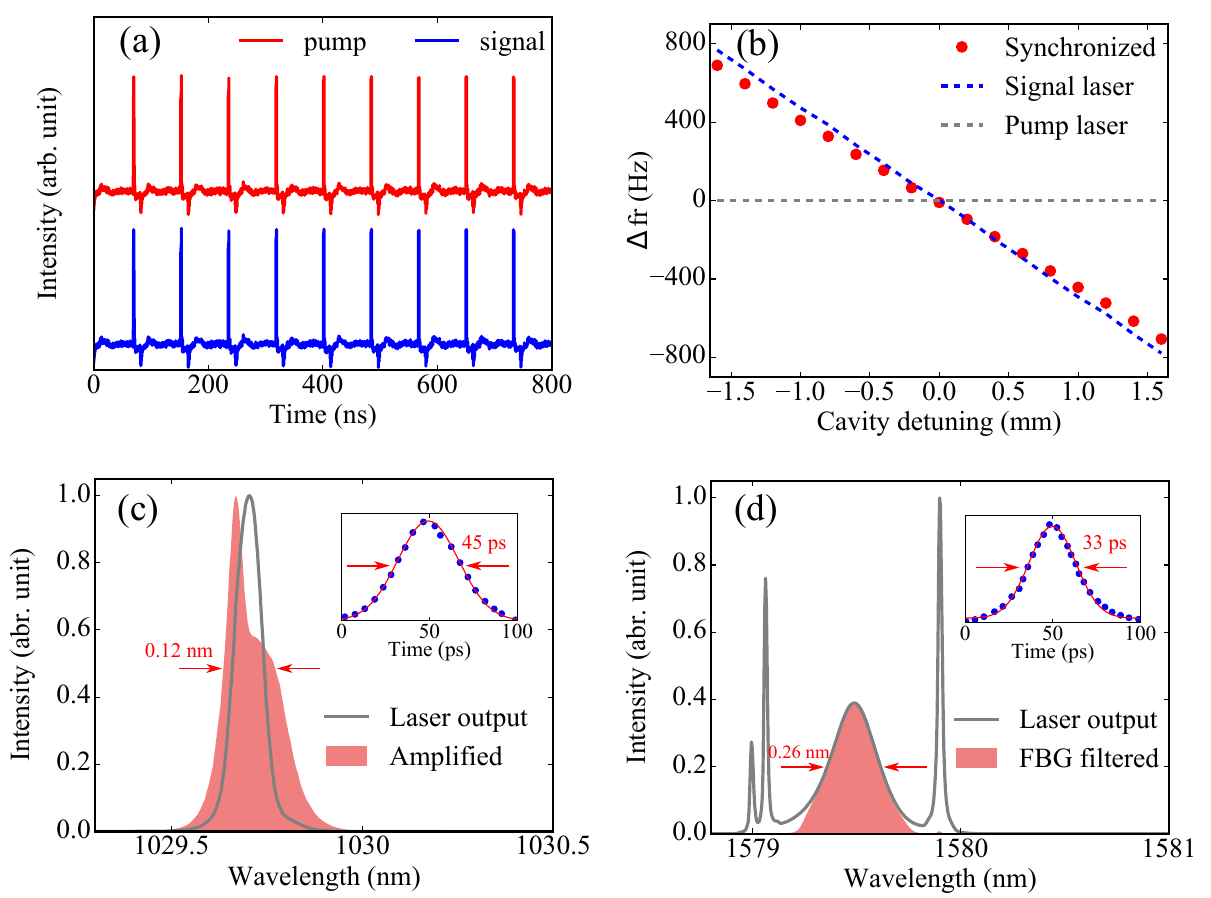}
\caption{(a) Pulse trains from synchronized pump and signal lasers. (b) Repetition rates of synchronized fiber lasers are kept the same as varying the cavity-length of signal laser. The dashed lines indicate the behavior of repetition rates for two free-running lasers. Note that the repetition rates are offset by the value of pump laser. Optical spectra of pump (c) and signal (d) fiber lasers. The area plots indicate the spectrum after power amplification for the pump and the spectrum after spectral filtering for the signal, respectively. The corresponding autocorrelation traces are given in the insets.} 
\label{fig2}
\end{figure}

Figure \ref{fig1}(a) gives the setup for the all-PM synchronization system, which consisted of an Yb-doped fiber laser (YDFL) and an Er-doped fiber laser (EDFL) in a shared-cavity configuration. Individually, the two fiber lasers could be passively mode-locked  at about 12.1 MHz by using the nonlinear amplifying loop mirror (NALM). Notably, a $\pi$/2 phase shifter (PS) was used in the phased-biased NALM to provide an additional phase difference, which could not only substantially reduce the mode-locking threshold, but also facilitate the self-initiation of mode-locked operation. The synchronous operation could be launched by simply adjusting the distance between the pair of collimators in the EDFL. The resulting synchronized pulse trains were shown in Fig. \ref{fig2}(a). The required mutual interaction for achieving the temporal synchronization was realized by nonlinear cross-phase modulation between the dual-color pulses within the common section of single-mode fiber. In contrast to previously reported passive synchronization technique \cite{Huang2012APL, Gu2010APL}, the all-polarization-maintaining design here enabled to launch the self-starting operation without the need of any polarization state optimization. Figure \ref{fig2}(b) presents the tight locking of the repetition rates for two synchronized fiber lasers as varying the cavity length of EDFL. The tolerance of the cavity-length mismatch was found to be about 3 mm, which corresponds to 1.5-kHz variation of the repetition rate. The achieved tolerance was more than three times longer than our previously reported synchronization system based on master-slave configuration \cite{Huang2018OE}. Moreover, the shared-cavity design here favored more compactness by eliminating the optical amplifier for the injected light.

\begin{figure}[t!]
\centering
\includegraphics[width=0.6\columnwidth]{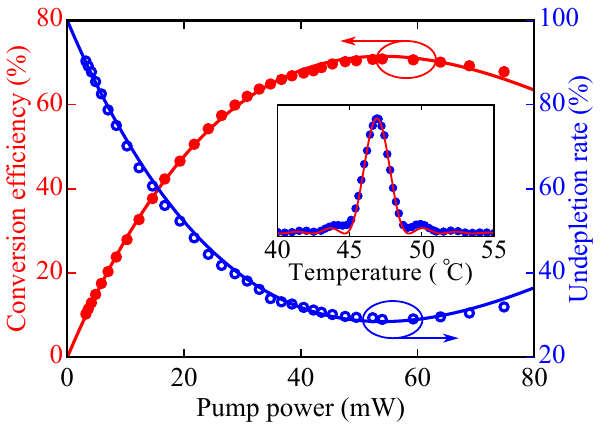}
\caption{Conversion efficiency and undepletion rate as a function of the pump power. The solid lines are calculated by the theoretical model given in the text. The inset shows the temperature dependence of conversion efficiency.}
\label{fig3}
\end{figure}

We now turn to characterize the output pulses from two fiber lasers. The spectrum from YDFL given in Fig. \ref{fig2}(c) was centered at 1029.7 nm with a full width at half maximum (FWHM) of 0.12 nm, which was mainly determined by the fiber Bragg grating (FBG) used in the laser cavity. As shown in Fig. \ref{fig1}(b), the output pulse from YDFL, after a fiber amplifier with bidirectional pumping, was used to provide the pump source in the subsequent nonlinear frequency conversion. The slightly broadened spectral width for the amplified pulses, as illustrated with the shaped area in Fig. \ref{fig2}(c), was due to non-negligible nonlinearity in the single-mode fiber. The corresponding autocorrelation trace was measured to be 45 ps, which indicated a pulse duration of 32 ps under the assumption of a Gaussian profile. In parallel, the EDFL delivered pulses at a central wavelength of 1579.5 nm as shown in Fig. \ref{fig2}(d). The exhibited spectral peaks were ascribed to characteristic Kelly sidebands for soliton mode-locked lasers \cite{Huang2018OE}. The output spectrum was further spectrally filtered by another FBG, hence resulting in a Gaussian-like profile with a FWHM bandwidth of 0.26 nm. The corresponding pulse duration was inferred to be 23 ps. Subsequently, the engineered pulses were used to provide the low-light-level signal source after intensive attenuation. It is worth noting that the temporal and spectral manipulation for the signal and pump sources was essential to optimize the performance of the upconversion detector to develop \cite{Gu2010APL}. At the power engineering stage, the attenuated signal and amplified pump pulses were temporally overlapped by adjusting the delay line as shown in Fig. \ref{fig1}, which were then spatially combined by a fiber wavelength division multiplexer (WDM) into a common single-mode fiber. Subsequently, the mixed beam was focused into a periodically-poled lithium niobate (PPLN) crystal with a length of 25 mm. As given in the inset of Fig. \ref{fig3}, quasi-phase matching was obtained at an operation temperature of 46.9 $^\circ$C  for a grating period of 11.49 $\mu$m. 

\section{Results and discussion}
We now proceed to investigate the performance of the implemented upconversion detector. In the experiment, the upconversion detector was firstly characterized in the low-light scenario. The signal average power was typically set at the level of 100 $\mu$W, which was about two orders of magnitude lower than that for the pump source. The involved power measurement could be conveniently conducted by a sensitive power meter. Therefore, the conversion efficiency could  be calculated from the measured power of the infrared signal and visible upconverted light, \textit{i.e.}, $P_\text{s}$ and $P_\text{up}$: $\eta = \frac{P_\text{up}}{P_\text{s}}  \times \frac{\lambda_\text{up}}{\lambda_\text{s}}\ ,$ where $\lambda_\text{s}$ and $\lambda_\text{up}$ denote the wavelengths for the signal and upconverted fields, respectively. Figure \ref{fig3} presents the inferred conversion efficiency as varying the pump power.  As expected, the conversion efficiency was observed to be saturated since further increasing the pumping would render the upconverted filed down-converted back to the original wavelength \cite{Xiang2018PRA}. In the non-depleted pump approximation, the conversion efficiency could be modeled with $\eta = \eta_\text{m} \sin^2(\sqrt{P/P_\text{m}} \pi/2) \ ,$ where $\eta_\text{m}$ is the expected maximum efficiency and $P_\text{m}$ is the corresponding pump power. $\eta_\text{m}$ and $P_\text{m}$ were fitted to be 72\% and 54 mW, respectively. Although complete frequency conversion could be obtained in principle, the limited efficiency here might be ascribed to the spatial-mode mismatching for the focused pump and signal beams within the nonlinear crystal.

Furthermore, the observed conversion efficiency was confirmed by using another technique, which relied on monitoring the power of the unconverted signal. The undepletion rate $R$ of signal field during the nonlinear conversion process can be defined by $R = 1 - \eta = P_\text{res}/ P_\text{0} \ ,$ where $P_\text{0}$ and $P_\text{res}$ are the input and residual signal power, respectively. In practice, the light power can be measured by a power meter or an optical detector. The advantage of the proposed method is the avoidance to calibrate power measurement devices for different wavelengths. Moreover, the conversion efficiency could be accurately inferred regardless of the total loss including propagation transmission and filtering efficiency, since the related optical power would be scaled with the same factor. As shown in Fig. \ref{fig1}(c), a flip mirror was used to redirect the output beam through two long-pass filters with a cut-off wavelength of 1200 nm. The measured data was presented with blue open circles in Fig. \ref{fig3}, which was modeled quite well with the previous obtained parameters. 

We now investigate the performance of the upconversion detector at the single-photon level. To this end, the signal source was intensively attenuated by a series of calibrated neutral density filters. The average photon number per pulse was set to be about 0.15. To reduce the background noise, the upconverted photons were steered through a filtering stage. As shown in Fig. \ref{fig1}(c), a short-pass filter with a cut-off wavelength of 950 nm and three band-pass filters with a bandwidth of 10 nm were combined to provide a noise rejection of 120 dB at the pump wavelength. The overall transmission of the filter set at the signal wavelength $\eta_\text{f}$ was measured to 84\%. The filtered field was subsequently coupled into a single-mode fiber with a coupling efficiency $\eta_\text{c}$ of 91\%, which could effectively isolate the ambient noise. Finally, the photons were detected by a fiber-coupled single-photon counting module (SPCM) with a detection efficiency $\eta_\text{SPCM}$ of 55\% and dark noise of 100 Hz.

\begin{figure}[b!]
\centering
\includegraphics[width=1\columnwidth]{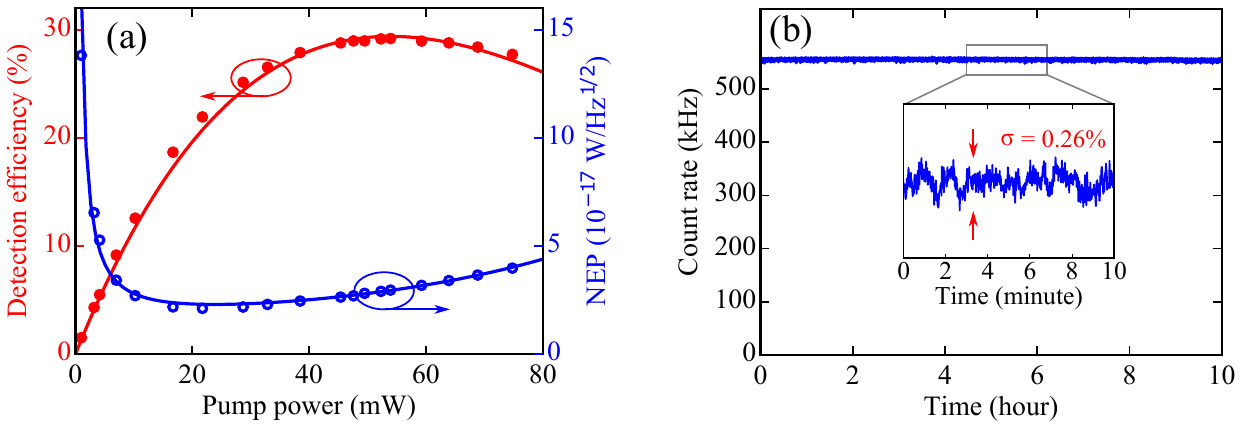}
\caption{(a) Detection efficiency and noise equivalent power (NEP) of the upconversion detector as a function of the pump power. The solid lines correspond to the theoretical model given in the text.  (b) Long-term stability of the count rate for the upconverted single-photon signal. $\sigma$ indicates the relative fluctuation.}
\label{fig4}
\end{figure}

At the peak conversion efficiency, the recorded maximum count rate was 550 kHz, which gave a detection efficiency $\eta_\text{d}$ about 30$\pm$1\% by considering the 1.815-MHz photon flux for the input infrared signal. By taking into account the total loss given by $\eta_\text{f} \times \eta_\text{c} \times \eta_\text{SPCM} = 0.42$, the conversion efficiency at the single-photon level was calculated to be 72$\pm$1\%, which was identical to the one obtained in the weak signal regime. The achieved detection efficiency was comparable to previous results with PPLN waveguide chips \cite{Ma2018JOSAB} and bulk crystals \cite{Gu2010APL}. Further increase of the detection efficiency to 50\% was possible with technological improvements in filtering stage, conversion unit and silicon detector, which may compete with the state-of-the-art gated InGaAs APDs \cite{Restelli2013APL, Tada2016CLEO}. It is worth noting that the proposed scheme would enable unique features by promoting immediate variations to realize photon-number-resolving detection and single-photon-level imaging, which may simulate broader applications at the telecom wavelengths.

As another important figure of merit, the background noise $N_\text{b}$ of the upconversion detector was measured to be 2.5 kHz at the peak detection efficiency, consequently led to a noise equivalent power (NEP = $h \nu \sqrt{2 N_\text{b}}/\eta_\text{d}$) of $3\times10^{-17}\ \text{W/Hz}^{1/2}$, where $h \nu$ denote the the energy of a signal photon. Figure \ref{fig4}(a) shows the detection efficiency and NEP as a function of the pump power. The lowest NEP was found to be $2\times10^{-17}\ \text{W/Hz}^{1/2}$ at a modest pump power of 22 mW, which corresponded to the highest signal-to-noise ratio of the upconversion photon detector. The pulse pump excitation used in our experiment confined the induced noise in a short time window, which significantly suppressed the noise counts. Additionally, the high spectral brightness of the narrow-bandwidth pump source could lower the required pump power for maximum nonlinear conversion, thus leading to further reduction of the background counts. It has been studied that the dominant sources of the background noise are upconverted spontaneous parametric downconversion (USPDC) and second-harmonic generation (SHG)-induced SPDC \cite{Meng2019OL}. In our experiment, the SHG-induced SPDC noise was centered at 578 nm, which was far away from the targeted SFG wavelength at 623 nm. Therefore, the residual background noise, that cannot be removed by our spectral filtering, was mainly from USPDC. To go beyond the achieved detection sensitivity, the use of long-wavelength-pumping technique might result in significant improvement \cite{Pelc2011OE}. Notably, our synchronization scheme can be readily modified by including a thulium-doped mode-locked fiber laser based on NALM \cite{Chernysheva2012PTL}, which could provide a longer-wavelength pump source at 2 $\mu$m. The efficacy of noise suppression has also been confirmed in the experiment by taking the YDFL output at 1.03 $\mu$m as the signal source. 

Finally, we have characterized the long-term stability of the upconverted signal as given in Fig. \ref{fig4}(b). A superior stability with a relative fluctuation of 0.26\% was exhibited over ten hours, which was at least one order of magnitude lower than previously reported results \cite{Pan2006OL}. It was the synergic result of stable fiber lasers and tight passive synchronization based on the all-polarization-maintaining configuration that enabled to achieve the aforementioned performance without any active measures for temperature stabilization and vibration isolation. Therefore, the demonstrated single-photon upconversion detector would be suitable for applications requiring long-time acquisition and stable photon counting.

It is worth noting that the used passive technique could enable to synchronize two ultrafast mode-locked fiber lasers with pulse widths of tens of femtoseconds. Therefore, the proof-of-principle demonstration of upconversion detection could be readily extended into the ultrafast regime, where ultrafast probes with a broadband spectrum would be used for instance in spectroscopic trace analysis. In this case, the ultrafast coincident pump would benefit high conversion efficiency and low background noise by confining the intensive excitation within a narrow time window.

\section{Conclusion}
In summary, we have experimentally implemented a coincidence-pumping upconversion detector based on passively synchronized all-PM fiber laser system. The internal conversion efficiency reached to 72\% owing to the spectro-temporal optimization of the signal and pump sources. Telecom photons were thus registered at a total detection efficiency of 30\% with a noise equivalent power of $3\times10^{-17}\ \text{W/Hz}^{1/2}$. Also, the demonstrated upconversion detector was featured with turn-key operation for the pump gating synchronization and long-term stability for the photon detection performance, which would be useful in ultra-sensitive pump-probe experiments and low-light active infrared sensing.

\end{document}